# Modeling, Prediction and Risk Management of Distribution System Voltages with Non-Gaussian Probability Distributions

Yuanhai Gao, Xiaoyuan Xu, Zheng Yan, Mohammad Shahidehpour, Bo Yang, and Xinping Guan

*Abstract*—High renewable energy penetration into power distribution systems causes a substantial risk of exceeding voltage security limits, which needs to be accurately assessed and properly managed. However, the existing methods usually rely on the joint probability models of power generation and loads provided by probabilistic prediction to quantify the voltage risks, where inaccurate prediction results could lead to over or under estimated risks. This paper proposes an uncertain voltage component (UVC) prediction method for assessing and managing voltage risks. First, we define the UVC to evaluate voltage variations caused by the uncertainties associated with power generation and loads. Second, we propose a Gaussian mixture model-based probabilistic UVC prediction method to depict the non-Gaussian distribution of voltage variations. Then, we derive the voltage risk indices, including value-at-risk (VaR) and conditional value-at-risk (CVaR), based on the probabilistic UVC prediction model. Third, we investigate the mechanism of UVC-based voltage risk management and establish the voltage risk management problems, which are reformulated into linear programming or mixed-integer linear programming for convenient solutions. The proposed method is tested on power distribution systems with actual photovoltaic power and load data and compared with those considering probabilistic prediction of nodal power injections. Numerical results show that the proposed method is computationally efficient in assessing voltage risks and outperforms existing methods in managing voltage risks. The deviation of voltage risks obtained by the proposed method is only 15% of that by the methods based on probabilistic prediction of nodal power injections.

## Nomenclature

### A. Indices and Sets

This work was supported in part by the National Natural Science Foundation of China under Grants U2166201 and 52077136 and Shanghai Science and Technology Plan Project under Grant 23DZ1201200. (Corresponding author: Xiaoyuan Xu)

Yuanhai Gao, Xiaoyuan Xu, and Zheng Yan are with the Ministry of Education, Key Laboratory of Control of Power Transmission and Conversion (Shanghai Jiao Tong University), Shanghai 200240, China, and also with the Shanghai Non-carbon Energy Conversion and Utilization Institute, Shanghai 200240, China (e-mail: gaoyuanhai@sjtu.edu.cn; xuxiaoyuan@sjtu.edu.cn; yanz@sjtu.edu.cn).

Mohammad Shahidehpour is with the Galvin Center for Electricity Innovation, Illinois Institute of Technology, Chicago, IL 60616 USA (e-mail: ms@iit.edu).

Bo Yang is with the Department of Automation and the Key Laboratory of System Control and Information Processing, Ministry of Education of China, Shanghai Jiao Tong University, Shanghai 200240, China, and also with the Shanghai Engineering Research Center of Industrial Intelligent Control and Management, Shanghai 200240, China (e-mail: bo.yang@sjtu.edu.cn).

Xinping Guan is with the Department of Automation and the Key Laboratory of System Control and Information Processing, Ministry of Education of China, Shanghai Jiao Tong University, Shanghai 200240, China (e-mail: xpguan@sjtu.edu.cn).

| | |
|---|---|
| $i$ | Index of buses |
| $j$ | Index of power providers |
| $g, d$ | Index of uncertain power generation and loads |
| $\bar{g}, \bar{d}$ | Index of constant power generation and loads |
| $\mathbb{I}$ | Set of buses excluding the slack bus |
| $\mathbb{J}$ | Set of power providers |
| $\mathbb{G}, \mathbb{D}$ | Set of uncertain power generation and loads |
| $\bar{\mathbb{G}}, \bar{\mathbb{D}}$ | Set of constant power generation and loads |

### B. Parameters

| | |
|---|---|
| $v_{i,\max}, v_{i,\min}$ | Maximum and minimum limits of squared voltage magnitude of bus $i$ |
| $\tau$ | Confidence level of voltage limits |
| $\boldsymbol{F}$ | Inverse matrix of a reduced branch-bus incidence matrix |
| $\boldsymbol{D}_\mathrm{r}, \boldsymbol{D}_\mathrm{x}$ | Diagonal matrices of branch resistance and reactance |
| $q_{j,\max}, q_{j,\min}$ | Maximum and minimum reactive power limits of provider $j$ |
| $c_j$ | Unit cost of purchasing reactive power from provider $j$ |
| $v_0$ | Squared voltage magnitude of the slack bus |
| $b$ | Sensitivity coefficient of squared voltage magnitude with respect to power generation and loads |
| $N$ | Sample number of uncertain power generation and loads |
| $L$ | Number of discrete realizations of $\alpha$ |

### C. Variables

| | |
|---|---|
| $\boldsymbol{P}, \boldsymbol{Q}$ | Vectors of active and reactive power injections |
| $v_i$ | Squared voltage magnitude of bus $i$ |
| $\boldsymbol{v}$ | Vector of $v_i$ |
| $v_i^\mathrm{r}, v_i^\mathrm{c}, v_i^\mathrm{o}$ | Uncertain, controllable, constant parts of $v_i$ |
| $\tilde{v}_i^\mathrm{r}$ | Prediction of $v_i^\mathrm{r}$ |
| $\boldsymbol{u}_i$ | Vector $[v_i^\mathrm{r}\ \tilde{v}_i^\mathrm{r}]^\mathrm{T}$ |
| $\chi_g, \zeta_d$ | Active power injections of uncertain power generation $g$ and load $d$ |
| $\tilde{\chi}_g, \tilde{\zeta}_d$ | Predictions of $\chi_g, \zeta_d$ |
| $\boldsymbol{\chi}, \boldsymbol{\zeta}$ | Vectors of $\chi_g, \zeta_d$ |
| $p_{\bar{g}}, p_{\bar{d}}$ | Active power of constant power generation $\bar{g}$ and load $\bar{d}$ |
| $p_j$ | Active power of provider $j$ |
| $q_j$ | Reactive power of provider $j$ |
| $q_j^*$ | Absolute value of $q_j$ |
| $\alpha$ | Active power curtailment ratio |

## I. INTRODUCTION

VOLTAGE management has played an important role in power distribution system operations due to the deep integration of renewable energy [1], [2], [3]. The power outputs of renewable energy power generation are variable [4]; thus, probability models have been utilized to fully characterize the uncertainties associated with renewable energy. In this context, voltage management aims to maintain the risk of violating voltage limits within a desired level in a cost-effective way [5].

The voltage risk is usually evaluated by the value-at-risk (VaR), and the corresponding voltage risk management problem is formulated as chance-constrained stochastic optimization [6], which is transformed into mixed-integer programming (MIP) via sample average approximation (SAA) for computational tractability [7]. The conditional value-at-risk (CVaR), which is a conservative approximation of VaR, is also a widely used risk index [8], and the optimization problem with CVaR constraints can also be solved in the SAA framework. However, the solution accuracy of SAA relies on the quantity and quality of samples. In addition, a large number of samples would cause heavy computational burdens.

To reduce the computational burdens of SAA methods in solving voltage risk management problems, some scholars have developed analytical expressions of VAR or CVaR constraints under specific probability models. In [9], wind power is modeled by Gaussian distributions, and the chance constraints of out-of-limits are transformed into cone constraints. In [10], the uncertain photovoltaic (PV) power generation is also modeled by Gaussian distributions, and the tractable counterparts of chance constraints are derived to solve the reactive power optimization problem. Ref. [11] extends the voltage risk management method in [10] to tackle the correlations among PV power. The Gaussian distribution-based probability models of renewable energy power generation allow for efficient risk calculations, but they cannot depict the non-Gaussian distributed uncertain factors.

The Gaussian mixture model (GMM) has been adopted to establish probability density distributions (PDFs) of renewable energy power generation [12], [13]. GMMs are presented as linear combinations of Gaussian distributions. Hence, they retain the merit of Gaussian distributions in deriving the formulas of risk indices while fitting complicated probability distributions of renewable energy [14]. In [15], the analytical expressions of voltage risks are deduced based on the GMM-based uncertainty models, and the voltage risk management problem is solved conveniently.

The aforementioned studies mainly focus on managing voltage risks with given probability models of nodal power injections. They have overlooked the changes in probability models over time. Compared with fixed probability models, it is more desirable to apply the probability models obtained by prediction techniques to power distribution system operations. To this end, in [16] and [17], the mixture density neural network is designed to predict the GMM-based marginal distribution of wind power, but the joint distribution is not modeled. In [18], the recurrent neural network is combined with low-rank Gaussian Copula to predict the joint PDF of renewable energy power generation. In [19], a multivariate probabilistic prediction model is designed based on monotone broad learning and Copula theory to predict the joint probability distribution of PV power. However, the joint probability distributions obtained by these multivariate probabilistic prediction methods are not in the GMM form, which causes the voltage risk management problems to only be solved by the time-consuming SAA techniques. Therefore, the traditional probabilistic prediction methods, which obtain the joint distributions of nodal power injections, do not support the efficient management of voltage risks.

This paper proposes a prediction method of uncertain voltage components (UVCs) for accurately assessing and efficiently managing voltage risks in power distribution system operations. A distinguishing feature of our study is that the univariate PDFs of each nodal voltage, rather than the joint PDF of nodal power injections, are predicted to manage voltage risks. The main contributions of this paper are summarized as follows:

1) The concept of UVCs is defined and derived from nodal voltages to capture the voltage variations caused by the exogenous uncertainties associated with power generation and loads, which allows for the effective management of voltage risks without resorting to the challenging task of conducting multivariate probabilistic predictions of nodal power injections.

2) The probabilistic prediction method of UVCs is proposed to depict the non-Gaussian distribution of voltage variations. Then, the computationally efficient formulas of VaR-based and CVaR-based voltage risk indices are established based on the probabilistic UVC prediction. In contrast, the existing methods usually use the time-consuming SAA techniques to calculate the risk indices of non-Gaussian distributed voltages.

3) The UVC-based voltage risk management problem is designed, and the computationally efficient solution method is developed. The proposed method is compared with the traditional voltage risk management methods which use the probabilistic prediction of nodal power injections. Results show that the deviation of voltage risks obtained by the proposed method is only 15% of that by other methods.

The rest of the paper is organized as follows: Section II gives the definition and prediction method of UVCs as well as the voltage risk assessment method. Section III discusses the mechanism of voltage risk management and gives the optimization problem and solution methods. Section IV compares the overall flowchart of the proposed method with that of the conventional method. Case studies are given in Section V, followed by the conclusions in Section VI.

## II. VOLTAGE RISK ASSESSMENT BASED ON UNCERTAIN VOLTAGE COMPONENT PREDICTION

In this section, the UVC is defined to depict the nodal voltage variations under nodal power injection uncertainties. Then, the probabilistic UVC prediction model is established. Finally, the voltage risk assessment method is presented.

### A. Definition of Uncertain Voltage Component

The squared nodal voltage magnitude of distribution systems can be approximated as linear functions of nodal power injections based on the linear DistFlow model [20]:

$$\upsilon_i = \sum_{\ell \in \mathbb{I}} \left( R_{i,\ell} P_\ell + X_{i,\ell} Q_\ell \right) + \upsilon_0 \quad (1)$$

where $P_\ell$ and $Q_\ell$ are the active and reactive power injections at bus $\ell$, respectively, $\upsilon_i$ is the squared voltage magnitude of bus $i$, $\upsilon_0$ is the squared voltage magnitude of the slack bus, and $R_{i,\ell}$ and $X_{i,\ell}$ are the sensitivity coefficients. The accuracy of this linear model in calculating voltages has been widely validated in the existing literature [1], [8], [10], [21], [22].

The nodal active power $P_\ell$ consists of uncertain power generation $\chi_g$, uncertain loads $\zeta_d$, constant power generation $p_{\bar{g}}$, constant loads $p_{\bar{d}}$, and controllable active power $p_j$. Without loss of generality, it is assumed that the nodal reactive power $Q_\ell$ consists of uncertain and constant power generation and loads with fixed power factors and controllable reactive power $q_j$. Hence, $P_\ell$ and $Q_\ell$ are expressed as:

$$\begin{cases} P_\ell = \sum_{g \in \mathbb{G}} \eta_{\ell,g} \chi_g + \sum_{\bar{g} \in \bar{\mathbb{G}}} \eta_{\ell,\bar{g}} p_{\bar{g}} - \sum_{d \in \mathbb{D}} \eta_{\ell,d} \zeta_d - \sum_{\bar{d} \in \bar{\mathbb{D}}} \eta_{\ell,\bar{d}} p_{\bar{d}} + \sum_{j \in \mathbb{J}} \eta_{\ell,j} p_j \\ Q_\ell = \sum_{g \in \mathbb{G}} \eta_{\ell,g} \kappa_g \chi_g + \sum_{\bar{g} \in \bar{\mathbb{G}}} \eta_{\ell,\bar{g}} \kappa_{\bar{g}} p_{\bar{g}} - \sum_{d \in \mathbb{D}} \eta_{\ell,d} \kappa_d \zeta_d \\ \quad - \sum_{\bar{d} \in \bar{\mathbb{D}}} \eta_{\ell,\bar{d}} \kappa_{\bar{d}} p_{\bar{d}} + \sum_{j \in \mathbb{J}} \eta_{\ell,j} q_j \end{cases} \quad (2)$$

where $\kappa$ is the tangent of power factor angles, and $\eta$ indicates the buses of power generation and loads. If the power generation or loads are connected to bus $\ell$, $\eta=1$; otherwise, $\eta=0$.

By submitting (2) into (1), the squared voltage magnitude $\upsilon_i$ is partitioned into three parts, stated as follows:

$$\upsilon_i = \upsilon_i^{\text{r}} + \upsilon_i^{\text{c}} + \upsilon_i^{\text{o}} \quad (3a)$$

$$\upsilon_i^{\text{r}} = \sum_{g \in \mathbb{G}} b_{i,g} \chi_g - \sum_{d \in \mathbb{D}} b_{i,d} \zeta_d \quad (3b)$$

$$\upsilon_i^{\text{c}} = \sum_{j \in \mathbb{J}} b_{i,j}^q q_j + \sum_{j \in \mathbb{J}} b_{i,j}^p p_j \quad (3c)$$

$$\upsilon_i^{\text{o}} = \sum_{\bar{g} \in \bar{\mathbb{G}}} b_{i,\bar{g}} p_{\bar{g}} - \sum_{\bar{d} \in \bar{\mathbb{D}}} b_{i,\bar{d}} p_{\bar{d}} + \upsilon_0 \quad (3d)$$

where $\upsilon_i^{\text{r}}$, $\upsilon_i^{\text{c}}$, and $\upsilon_i^{\text{o}}$ are defined as the uncertain, controllable, and constant voltage components, respectively, and $b$ represents the sensitivity coefficients of squared voltage magnitude with respect to power generation and loads.

The UVC $\upsilon_i^{\text{r}}$ is determined by the exogenous uncertainties associated with power generation $\chi_g$ and loads $\zeta_d$. The controllable voltage component $\upsilon_i^{\text{c}}$ is determined by the controllable power $p_j$ and $q_j$, which are provided by power generation units, reactive power compensators, and other controllable resources. The constant voltage component $\upsilon_i^{\text{o}}$ is derived from the slack bus's voltage $\upsilon_0$ and the constant power generation $p_{\bar{g}}$ and loads $p_{\bar{d}}$.

*B. Probabilistic Prediction of Uncertain Voltage Component*

According to (3a), UVCs cause voltage uncertainties. Hence, the probabilistic UVC prediction method is developed in this subsection to evaluate voltage risks. To be specific, a joint probability distribution of true and predicted UVCs is established to perform probabilistic prediction.

*1) Building Joint Distributions of True and Predicted UVCs:*
**Step 1.** Train prediction models of power generation and loads

The probabilistic UVC prediction model is constructed on the point prediction models of power generation and loads. In this study, we employ the widely used time-series prediction model, long short-term memory network (LSTM), to establish the point prediction models of power generation and loads. Let $\chi_{g,n}^{\text{hist}}$ and $\zeta_{d,n}^{\text{hist}}$ denote the $n_{\text{th}}$ historical samples of power generation and loads, respectively, and let $\tilde{\chi}_{g,n}^{\text{hist}}$ and $\tilde{\zeta}_{d,n}^{\text{hist}}$ be their historical predictions.

**Step 2.** Acquire historical samples of true and predicted UVCs

Let $\upsilon_i^{\text{r}}$ and $\tilde{\upsilon}_i^{\text{r}}$ denote UVCs under true and predicted power generation and loads, respectively, and their samples are obtained by:

$$\begin{cases} V_{i,n} = \sum_{g \in \mathbb{G}} b_{i,g} \chi_{g,n}^{\text{hist}} - \sum_{d \in \mathbb{D}} b_{i,d} \zeta_{d,n}^{\text{hist}}, \quad n \in \{1,2,\cdots N\} \\ \tilde{V}_{i,n} = \sum_{g \in \mathbb{G}} b_{i,g} \tilde{\chi}_{g,n}^{\text{hist}} - \sum_{d \in \mathbb{D}} b_{i,d} \tilde{\zeta}_{d,n}^{\text{hist}}, \quad n \in \{1,2,\cdots N\} \end{cases} \quad (4)$$

where $V_{i,n}$ and $\tilde{V}_{i,n}$ are the $n_{\text{th}}$ samples of $\upsilon_i^{\text{r}}$ and $\tilde{\upsilon}_i^{\text{r}}$, respectively, and $N$ is the number of samples.

**Step 3.** Establish the joint probability distribution of true and predicted UVCs

The joint probability distribution of true and predicted UVCs is established by kernel density estimation (KDE) as:

$$f^{(\text{b})}(\boldsymbol{u}_i) = \sum_{n=1}^{N} \frac{1}{N} \text{N}(\boldsymbol{u}_i | \boldsymbol{U}_{i,n}, \boldsymbol{H}_{i,n}) \quad (5)$$

where $\text{N}(\cdot)$ is the Gaussian kernel function, $\boldsymbol{u}_i$ represents vector $[\upsilon_i^{\text{r}} \ \tilde{\upsilon}_i^{\text{r}}]^{\text{T}}$, $\boldsymbol{U}_{i,n}$ is the $n_{\text{th}}$ sample of $\boldsymbol{u}_i$ and $\boldsymbol{U}_{i,n} = [V_{i,n} \ \tilde{V}_{i,n}]^{\text{T}}$, and $\boldsymbol{H}_{i,n}$ is the bandwidth matrix at $\boldsymbol{U}_{i,n}$ and is determined by Silverman's rule of thumb, as stated below:

$$\begin{cases} \boldsymbol{H}_{i,n} = \text{diag}([h_{i,n}^2, \tilde{h}_{i,n}^2]^{\text{T}}) \\ h_{i,n} = 0.9 \min(\sigma_i, I_{QR,i}/1.34) N^{-0.2} \\ \tilde{h}_{i,n} = 0.9 \min(\tilde{\sigma}_i, \tilde{I}_{QR,i}/1.34) N^{-0.2} \end{cases} \quad (6)$$

where $\text{diag}(\cdot)$ is the operator to transform a vector to a diagonal matrix, $\sigma_i$ and $\tilde{\sigma}_i$ are the standard deviations of samples $\{V_{i,1}, \cdots V_{i,N}\}$ and $\{\tilde{V}_{i,1}, \cdots \tilde{V}_{i,N}\}$, respectively, and $I_{QR,i}$ and $\tilde{I}_{QR,i}$ are the interquartile ranges of samples $\{V_{i,1}, \cdots V_{i,N}\}$ and $\{\tilde{V}_{i,1}, \cdots \tilde{V}_{i,N}\}$, respectively.

Considering that Gaussian-based KDE consists of Gaussian distributions, $f^{(\text{b})}(\boldsymbol{u}_i)$ is presented as the following GMM [23]:

$$f^{(\text{b})}(\boldsymbol{u}_i) \triangleq \sum_{k=1}^{K_{\text{b}}} \omega_{i,k}^{(\text{b})} \text{N}(\boldsymbol{u}_i | \boldsymbol{\mu}_{i,k}^{(\text{b})}, \boldsymbol{\Sigma}_{i,k}^{(\text{b})}) \quad (7)$$

where $K_{\text{b}}$ is the number of Gaussian components and equals the number of samples $N$. For the $k_{\text{th}}$ Gaussian component, the covariance matrix $\boldsymbol{\Sigma}_{i,k}^{(\text{b})}$ is $\boldsymbol{H}_{i,k}$, the mean vector $\boldsymbol{\mu}_{i,k}^{(\text{b})}$ is $\boldsymbol{U}_{i,k}$, and the weight coefficient $\omega_{i,k}^{(\text{b})}$ is $1/N$.

Although equation (7) will provide analytical formulas for voltage risks, the large number of Gaussian components of $f^{(\text{b})}(\boldsymbol{u}_i)$ leads to heavy computational burdens. Hence, the density-preserving hierarchical expectation maximum (DPHEM) algorithm [23] is employed to generate a new GMM $f(\boldsymbol{u}_i)$ with a smaller component number while retaining the key statistical feature of $f^{(\text{b})}(\boldsymbol{u}_i)$:

$$f(\boldsymbol{u}_i) = \sum_{k=1}^{K} \omega_{i,k} \text{N}(\boldsymbol{u}_i | \boldsymbol{\mu}_{i,k}, \boldsymbol{\Sigma}_{i,k}) \quad (8a)$$

$$\boldsymbol{\mu}_{i,k} = [\mu_{i,k}^{\Re}, \mu_{i,k}^{\text{R}}]^{T} \quad (8b)$$

$$\boldsymbol{\Sigma}_{i,k} = \begin{bmatrix} \Sigma_{i,k}^{\Re} & \Sigma_{i,k}^{\Re R} \\ \Sigma_{i,k}^{R\Re} & \Sigma_{i,k}^{R} \end{bmatrix} \quad (8c)$$

where $\boldsymbol{\Sigma}_{i,k}$, $\boldsymbol{\mu}_{i,k}$, and $\omega_{i,k}$ are the covariance matrix, mean vector, and weight coefficient of the $k_{\text{th}}$ Gaussian component, respectively, and $K$ is the number of Gaussian components.

*2) Performing Probabilistic UVC Prediction:*
**Step 1.** Predict power generation and loads

The predictions of power generation and loads are denoted as $\tilde{\chi}_g$ and $\tilde{\zeta}_d$, respectively.

**Step 2.** Predict UVCs

The prediction of UVC $\tilde{\upsilon}_i^{\text{r}}$ is obtained as:

$$\tilde{\upsilon}_i^{\text{r}} = \sum_{g \in \mathbb{G}} b_{i,g} \tilde{\chi}_g - \sum_{d \in \mathbb{D}} b_{i,d} \tilde{\zeta}_d \quad (9)$$

**Step 3.** Obtain probability distributions of UVCs

Based on the joint PDF of $\upsilon_i^{\text{r}}$ and $\tilde{\upsilon}_i^{\text{r}}$ (8) and the law of total probability, the conditional PDF of $\upsilon_i^{\text{r}}$ with given $\tilde{\upsilon}_i^{\text{r}}$ is obtained as:

$$f\left(\upsilon_i^{\text{r}} | \tilde{\upsilon}_i^{\text{r}}\right) = \frac{f(\boldsymbol{u}_i)}{f(\tilde{\upsilon}_i^{\text{r}})} = \sum_{k=1}^{K} \omega_{i,k}^* \, \text{N}\left(\upsilon_i^{\text{r}} | \mu_{i,k}^*, \Sigma_{i,k}^*\right) \quad (10a)$$

$$\mu_{i,k}^* = \mu_{i,k}^{\text{R}} + \Sigma_{i,k}^{\text{R}\Re} \left(\Sigma_{i,k}^{\Re}\right)^{-1} \left(\tilde{\upsilon}_i^{\text{r}} - \mu_{i,k}^{\Re}\right) \quad (10b)$$

$$\Sigma_{i,k}^* = \Sigma_{i,k}^{\text{R}} - \Sigma_{i,k}^{\text{R}\Re} \left(\Sigma_{i,k}^{\Re}\right)^{-1} \Sigma_{i,k}^{\Re \text{R}} \quad (10c)$$

$$\omega_{i,k}^* = \frac{\omega_{i,k} \, \text{N}\left(\tilde{\upsilon}_i^{\text{r}} | \mu_{i,k}^{\Re}, \Sigma_{i,k}^{\Re}\right)}{\sum_{k'=1}^{K} \omega_{i,k'} \, \text{N}\left(\tilde{\upsilon}_i^{\text{r}} | \mu_{i,k'}^{\Re}, \Sigma_{i,k'}^{\Re}\right)} \quad (10d)$$

where $\Sigma_{i,k}^*$, $\mu_{i,k}^*$, and $\omega_{i,k}^*$ are the variance, mean, and weight coefficient of the $k_{\text{th}}$ Gaussian component of the conditional PDF, respectively.

Fig. 1 gives the probabilistic prediction flowchart.

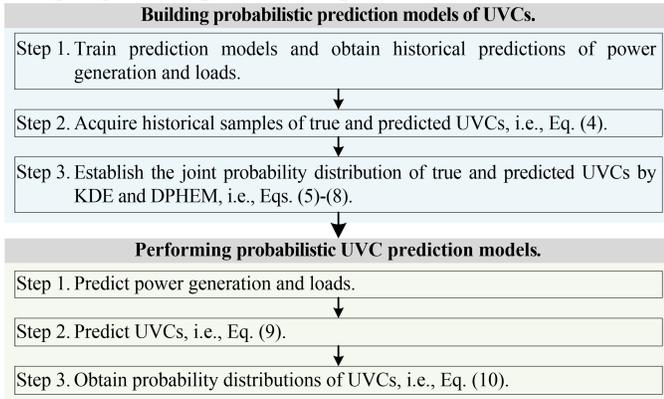

Fig. 1. Flowchart of probabilistic prediction model building and applications.

**Remark 1.** The constant coefficients $b_{i,g}$ and $b_{i,d}$ in (9) are derived from the linear DistFlow model; hence, the uncertainties of the prediction error of UVCs are entirely caused by the uncertainties associated with power generation and loads.

*C. Voltage Risk Assessment*

The VaR and CVaR are two widely used voltage risk indices, which are defined as follows:

$$\text{VaR}_\tau(\upsilon_i) \triangleq F_{\upsilon_i}^{-1}(\tau) \quad (11)$$

$$\text{CVaR}_\tau(\upsilon_i) \triangleq \frac{1}{1-\tau} \int_{\text{VaR}_\tau(\upsilon_i)}^{+\infty} x f_{\upsilon_i}(x) dx \quad (12)$$

where $F_{\upsilon_i}(\cdot)$ and $f_{\upsilon_i}(\cdot)$ are the cumulative distribution function (CDF) and PDF of variable $\upsilon_i$, respectively, and $1-\tau$ represents the risk level. Equation (11) indicates that $\text{Pr}\{\upsilon_i \leqslant \text{VaR}_\tau(\upsilon_i)\} = \tau$.

By replacing $\upsilon_i$ with its uncertain, controllable, and constant voltage components, $\text{VaR}_\tau(\upsilon_i)$ is expressed as

$$\text{VaR}_\tau(\upsilon_i) = \text{VaR}_\tau(\upsilon_i^{\text{r}} | \tilde{\upsilon}_i^{\text{r}}) + \upsilon_i^{\text{c}} + \upsilon_i^{\text{o}} \quad (13)$$

where $\tilde{\upsilon}_i^{\text{r}}$ is determined by the predictions of power generation and loads, and $\upsilon_i^{\text{r}}$ is a random variable. The derivation procedure of (13) is given in Appendix A.

$\text{VaR}_\tau(\upsilon_i^{\text{r}} | \tilde{\upsilon}_i^{\text{r}})$ is obtained by solving the equation:

$$F(\upsilon_i^{\text{r}} | \tilde{\upsilon}_i^{\text{r}}) = \tau \quad (14)$$

where $F(\upsilon_i^{\text{r}} | \tilde{\upsilon}_i^{\text{r}})$ is the conditional CDF, which is obtained from (10). Here, the nonlinear equation (14) is solved by the Newton-Raphson method, and the calculation method is given in Appendix B.

Similarly, $\text{CVaR}_\tau(\upsilon_i)$ is expressed as

$$\text{CVaR}_\tau(\upsilon_i) = \text{CVaR}_\tau(\upsilon_i^{\text{r}} | \tilde{\upsilon}_i^{\text{r}}) + \upsilon_i^{\text{c}} + \upsilon_i^{\text{o}} \quad (15)$$

where the closed-form expression of $\text{CVaR}_\tau(\upsilon_i^{\text{r}} | \tilde{\upsilon}_i^{\text{r}})$ is given as:

$$\text{CVaR}_\tau(\upsilon_i^{\text{r}} | \tilde{\upsilon}_i^{\text{r}}) = \frac{1}{1-\tau} \sum_{k=1}^{K} \omega_{i,k}^* \begin{pmatrix} \mu_{i,k}^* \left\{1 - F_k\left[\text{VaR}_\tau(\upsilon_i^{\text{r}} | \tilde{\upsilon}_i^{\text{r}})\right]\right\} \\ + \Sigma_{i,k}^* f_k\left[\text{VaR}_\tau(\upsilon_i^{\text{r}} | \tilde{\upsilon}_i^{\text{r}})\right] \end{pmatrix} \quad (16)$$

where $f_k[\cdot]$ and $F_k[\cdot]$ are the PDF and CDF of the $k_{\text{th}}$ Gaussian component of (10a), respectively. The derivation procedures of (15) and (16) are given in Appendices C and D, respectively.

## III. Voltage Risk Management Based on Uncertain Voltage Component Prediction

In this section, the mechanism of voltage risk management is investigated. Then, the voltage risk management problem is established and the solution method is given.

*A. Mechanism of Voltage Risk Management*

The following chance constraints are designed to manage the risks of violating voltage limits within a desired level:

$$\text{Pr}\{\upsilon_i > \upsilon_{i,\max}\} \leq 1-\tau, \quad \forall i \in \mathbb{I} \quad (17a)$$

$$\text{Pr}\{\upsilon_i < \upsilon_{i,\min}\} \leq 1-\tau, \quad \forall i \in \mathbb{I} \quad (17b)$$

where $\text{Pr}\{\cdot\}$ is the probability operator. According to the VaR definition, $\text{VaR}_\tau(\upsilon_i)$ and $-\text{VaR}_\tau(-\upsilon_i)$ indicate that $\text{Pr}\{\upsilon_i \leqslant \text{VaR}_\tau(\upsilon_i)\} = \tau$ and $\text{Pr}\{\upsilon_i \geqslant -\text{VaR}_\tau(-\upsilon_i)\} = \tau$, respectively; thus, (17) is equivalently presented as

$$\text{VaR}_\tau(\upsilon_i) \leq \upsilon_{i,\max}, \quad \forall i \in \mathbb{I} \quad (18a)$$

$$-\text{VaR}_\tau(-\upsilon_i) \geq \upsilon_{i,\min}, \quad \forall i \in \mathbb{I} \quad (18b)$$

Substituting (13) into (18), we obtain

$$\text{VaR}_\tau(\upsilon_i^{\text{r}} | \tilde{\upsilon}_i^{\text{r}}) + \upsilon_i^{\text{c}} + \upsilon_i^{\text{o}} \leq \upsilon_{i,\max}, \quad \forall i \in \mathbb{I} \quad (19a)$$

$$-\text{VaR}_\tau(-\upsilon_i^{\text{r}} | \tilde{\upsilon}_i^{\text{r}}) + \upsilon_i^{\text{c}} + \upsilon_i^{\text{o}} \geq \upsilon_{i,\min}, \quad \forall i \in \mathbb{I} \quad (19b)$$

There are two means to manage the voltage risks to satisfy (19), i.e., changing controllable voltage components $\upsilon_i^{\text{c}}$ and UVCs $\upsilon_i^{\text{r}}$, as discussed below.

*1) Changing Controllable Voltage Components*:

As shown in Fig. 2, the initial $\text{VaR}_\tau(v_i)$ is higher than $v_{i,\max}$, indicating the violation of constraint (19a). Then, the controllable voltage component $v_i^c$ is changed to make $\text{VaR}_\tau(v_i)$ and $-\text{VaR}_\tau(-v_i)$ fall within $[v_{i,\min}, v_{i,\max}]$, while the variation range of UVCs, indicated by $\text{VaR}_\tau(v_i^r | \tilde{v}_i^r)$ and $-\text{VaR}_\tau(-v_i^r | \tilde{v}_i^r)$, remains constant.

*2) Changing Uncertain Voltage Components:*

When the distance between the upper and lower bounds of UVCs is larger than $v_{i,\max} - v_{i,\min}$ ($\text{VaR}_\tau(v_i^r | \tilde{v}_i^r) + \text{VaR}_\tau(-v_i^r | \tilde{v}_i^r) > v_{i,\max} - v_{i,\min}$), constraint (19) cannot be satisfied by only changing the controllable voltage component $v_i^c$, as shown in the infeasible case of Fig. 3. In this case, the active power of renewable energy generation needs to be curtailed to reduce the UVC variations. Let $\alpha$ be the curtailment ratio of active power, which is set as the same for each renewable energy unit for the sake of fairness. The UVC in (3b) is modified as:

$$v_i^r = (1-\alpha)\sum_{g\in\mathbb{G}} b_{i,g}\chi_g - \sum_{d\in\mathbb{D}} b_{i,d}\zeta_d \quad (20)$$

According to (20), the active power curtailment restricts the UVC variations. Constraint (19) would be satisfied when the distance between the upper and lower bounds of voltages is no larger than $v_{i,\max} - v_{i,\min}$ ($\text{VaR}_\tau(v_i^r | \tilde{v}_i^r) + \text{VaR}_\tau(-v_i^r | \tilde{v}_i^r) \leq v_{i,\max} - v_{i,\min}$), as shown in the feasible case of Fig. 3.

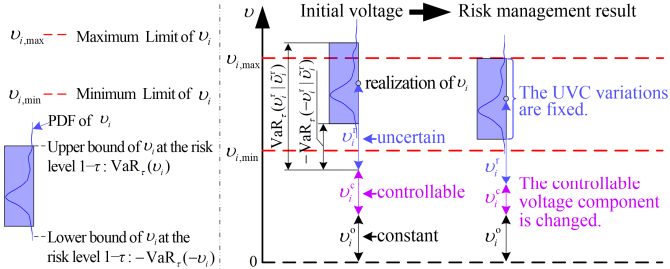

Fig. 2. Explanation of changing controllable voltage components.

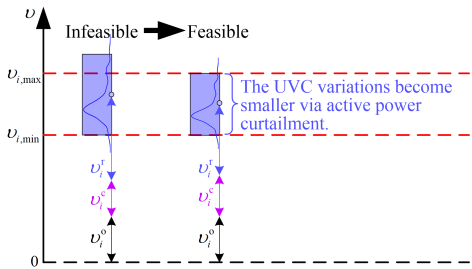

Fig. 3. Explanation of changing uncertain voltage components.

### B. Optimization Model of Voltage Risk Management

Reactive power is an important and economic resource to adjust voltages [24] for maintaining desired voltage risk levels, and its regulation can be regarded as an ancillary service in distribution system operations [1], [25]. In this context, the voltage risk management problem is formulated as:

$$\min_{q_j^*, q_j} \sum_{j\in\mathbb{J}} c_j q_j^* \quad (21a)$$

s.t.
$$q_j^* \geq q_j, \quad \forall j \in \mathbb{J} \quad (21b)$$
$$q_j^* \geq -q_j, \quad \forall j \in \mathbb{J} \quad (21c)$$
$$q_j \leq q_{j,\max}, \quad \forall j \in \mathbb{J} \quad (21d)$$
$$q_j \geq q_{j,\min}, \quad \forall j \in \mathbb{J} \quad (21e)$$
$$\Pr\{v_i > v_{i,\max}\} \leq 1-\tau, \quad \forall i \in \mathbb{I} \quad (21f)$$
$$\Pr\{v_i < v_{i,\min}\} \leq 1-\tau, \quad \forall i \in \mathbb{I} \quad (21g)$$
$$v = 2FD_r F^T P + 2FD_x F^T Q + v_0 \mathbf{1} \quad (21h)$$

where $P$ and $Q$ are the vectors of nodal active and reactive power injections, respectively, and $\mathbf{1}$ is a vector of all ones. The objective function of the problem is to minimize the cost of purchasing reactive power. Constraints (21b) and (21c) indicate the absolute value of $q_j$. Constraints (21d) and (21e) are the reactive power limits of reactive power provider $j$. Constraints (21f) and (21g) are the voltage risk constraints. Constraint (21h) represents the linear DistFlow model [20].

*1) VaR-based Reformulation Model:*

Based on (3) and (19), problem (21) is reformulated into an LP problem as:

$$\min_{q_j^*, q_j, v_i^c} \sum_{j\in\mathbb{J}} c_j q_j^* \quad (22a)$$

s.t.
$$v_i^c \leq v_{i,\max} - v_i^o - \text{VaR}_\tau(v_i^r | \tilde{v}_i^r), \forall i \in \mathbb{I} \quad (22b)$$
$$-v_i^c \leq -v_{i,\min} + v_i^o - \text{VaR}_\tau(-v_i^r | \tilde{v}_i^r), \quad \forall i \in \mathbb{I} \quad (22c)$$
$$v_i^c = \sum_{j\in\mathbb{J}} b_{i,j}^q q_j \quad (22d)$$
$$v_i^o = \sum_{\bar{g}\in\bar{\mathbb{G}}} b_{i,\bar{g}} p_{\bar{g}} - \sum_{\bar{d}\in\bar{\mathbb{D}}} b_{i,\bar{d}} p_{\bar{d}} + v_0 \quad (22e)$$
and (21b)-(21e)

where the decision variables include controllable voltage components $v_i^c$ and reactive power $q_j$. $\text{VaR}_\tau(v_i^r | \tilde{v}_i^r)$ and $\text{VaR}_\tau(-v_i^r | \tilde{v}_i^r)$ are obtained by solving (14).

*2) CVaR-based Reformulation Model:*

The VaR constraints (22b) and (22c) are conservatively replaced with CVaRs. Then, problem (22) is converted into:

$$\min_{q_j^*, q_j, v_i^c} \sum_{j\in\mathbb{J}} c_j q_j^* \quad (23a)$$

s.t.
$$v_i^c \leq v_{i,\max} - v_i^o - \text{CVaR}_\tau(v_i^r | \tilde{v}_i^r), \quad \forall i \in \mathbb{I} \quad (23b)$$
$$-v_i^c \leq -v_{i,\min} + v_i^o - \text{CVaR}_\tau(-v_i^r | \tilde{v}_i^r), \quad \forall i \in \mathbb{I} \quad (23c)$$
and (21b)-(21e), (22d)-(22e).

where $\text{CVaR}_\tau(v_i^r | \tilde{v}_i^r)$ and $\text{CVaR}_\tau(-v_i^r | \tilde{v}_i^r)$ are calculated by (16).

*3) Extended Model Considering Active Power Curtailment:*

According to (20), when the active power curtailment is considered to manage voltage risks, UVCs become decision-dependent uncertainties, which rely on the curtailment ratio of active power $\alpha$ [26]. Therefore, the risk indices $\text{VaR}_\tau(v_i^r | \tilde{v}_i^r)$, $\text{VaR}_\tau(-v_i^r | \tilde{v}_i^r)$, $\text{CVaR}_\tau(v_i^r | \tilde{v}_i^r)$, and $\text{CVaR}_\tau(-v_i^r | \tilde{v}_i^r)$ become functions of $\alpha$, rather than constants. To solve the voltage risk management problem, the risk indices are approximated as piecewise linearization functions of $\alpha$, which is presented as follows:

Let $\{\alpha_1, \ldots, \alpha_L\}$ be discrete realizations of $\alpha$. For each realization in $\{\alpha_1, \ldots, \alpha_L\}$, $\text{VaR}_\tau(v_i^r | \tilde{v}_i^r)$ and $\text{VaR}_\tau(-v_i^r | \tilde{v}_i^r)$ are obtained by the method in Section II and are denoted as

{$\beta_{i,1}$, …, $\beta_{i,L}$} and {$\gamma_{i,1}$, …, $\gamma_{i,L}$}, respectively. Then, $\text{VaR}_\tau(\upsilon_i^r | \tilde{\upsilon}_i^r)$ and $\text{VaR}_\tau(-\upsilon_i^r | \tilde{\upsilon}_i^r)$ are approximated as:

$$\begin{cases} \text{VaR}_\tau(\upsilon_i^r | \tilde{\upsilon}_i^r) = \sum_{l=1}^L \lambda_l \beta_{i,l} \\ \text{VaR}_\tau(-\upsilon_i^r | \tilde{\upsilon}_i^r) = \sum_{l=1}^L \lambda_l \gamma_{i,l} \\ \alpha = \sum_{l=1}^L \lambda_l \alpha_{i,l} \end{cases} \quad (24)$$

where $\lambda_l \geq 0$, $\sum_{l=1}^L \lambda_l = 1$, and {$\lambda_1$, …, $\lambda_L$} $\in \mathbb{SOS}_2$ (special ordered set of type 2). The $\mathbb{SOS}_2$ variables are formulated using auxiliary integer variables and linear inequality constraints. The detailed expressions are found in [27].

Finally, problem (22) is converted into the following mixed-integer linear programming (MILP) problem:

$$\min_{q_j^*, q_j, \upsilon_i^c, \alpha, \lambda_l} \sum_{j \in \mathbb{J}} c_j q_j^* + M\alpha \quad (25a)$$

s.t.
$$\sum_{l=1}^L \lambda_l \beta_{i,l} + \upsilon_i^c \leq \upsilon_{i,\max} - \upsilon_i^o, \quad \forall i \in \mathbb{I} \quad (25b)$$

$$\sum_{l=1}^L \lambda_l \gamma_{i,l} - \upsilon_i^c \leq -\upsilon_{i,\min} + \upsilon_i^o, \quad \forall i \in \mathbb{I} \quad (25c)$$

$$\lambda_l \geq 0, \quad \forall l \in \{1, 2, \cdots, L\} \quad (25e)$$

$$\sum_{l=1}^L \lambda_l = 1 \quad (25f)$$

$$\{\lambda_1, ..., \lambda_L\} \in \mathbb{SOS}_2 \quad (25g)$$

$$\alpha = \sum_{l=1}^L \lambda_l \alpha_{i,l} \quad (25h)$$

and (21b)-(21e), (22d)-(22e).

The active power is curtailed only when the risk constraints in (22) cannot be satisfied; thus, a large positive number $M$ is used as the weight of $\alpha$ in (25).

**Remark 2.** The CVaR is also approximated as a piecewise linear function of $\alpha$, and the CVaR-based voltage risk management problem is presented in a similar form as (25).

IV. OVERALL PROCEDURE

Fig.4 gives the overall flowchart of the proposed voltage risk assessment and management method based on UVC prediction (abbreviated as the UVCP method). In addition, the flowchart of the conventional method, which combines probabilistic prediction of nodal power injections with optimization methods (abbreviated as the PPO method), is also shown for comparison. In the PPO method, the probability distributions of power generation and loads are predicted, and then the voltage risk assessment and management problems are solved with the uncertainty models of nodal power injections. The optimization model of the PPO method is given in Appendix E. The key distinction between the proposed UVCP method and the conventional PPO method is the difference in probabilistic prediction targets in steps 1 and 2. The PPO method provides analytical expressions of voltage risks in step 4 using joint Gaussian distributions of nodal power injections, thus reducing the difficulty of solving the voltage risk management problem in steps 5 and 6. However, the PPO method cannot tackle the voltage variations caused by non-Gaussian uncertainties. In contrast, the UVCP method accurately captures these variations while supporting the analytical assessment and efficient management of voltage risks.

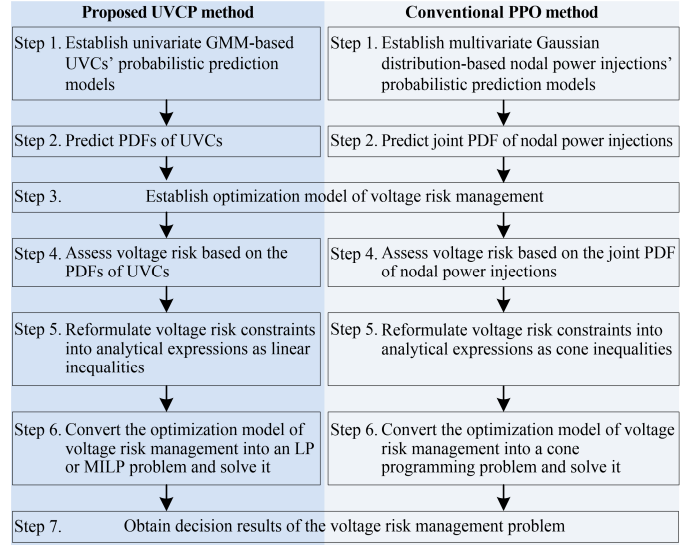

Fig. 4. Flowchart of voltage risk assessment and management by the proposed UVCP method and the conventional PPO method.

V. CASE STUDIES

*A. Simulation Setup*

The IEEE 33-bus [28] and 123-bus [29] distribution systems are used to test the performance of the proposed voltage risk assessment and management method based on UVCP method. The 33-bus distribution system is augmented with four PV units, each of which has a rated capacity of 1 MW. The 123-bus distribution system is augmented with nine PV units, each of which has a rated capacity of 0.8 MW. Figs. 5 and 6 show the installed buses of PV units. Table I gives the regulation range of reactive power. The unit cost $c_j$ of purchasing reactive power is set as \$20/Mvar. The voltage magnitude of the bus controlled by the upstream grid is set as 1.0 pu. The maximum and minimum voltage magnitudes are 1.05 pu and 0.95 pu, respectively. The desired risk level of voltage violation is set as 0.05.

The active power outputs of PV units are obtained using the actual solar irradiation data provided by SolarAnywhere [30]. The load data are generated by scaling the original loads of two systems [28], [29] according to the real-world load curves [31]. The hourly data from Jan. 1, 2011 to Dec. 31, 2014 are used in the case studies. The first 70% of the data is used as the training data to obtain the prediction models of UVCs, the rest is used as the testing data. These power generation and load data have been uploaded to the GitHub website [32].

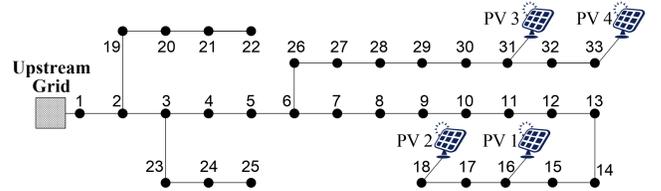

Fig. 5. IEEE 33-bus system augmented with 4 PV units.

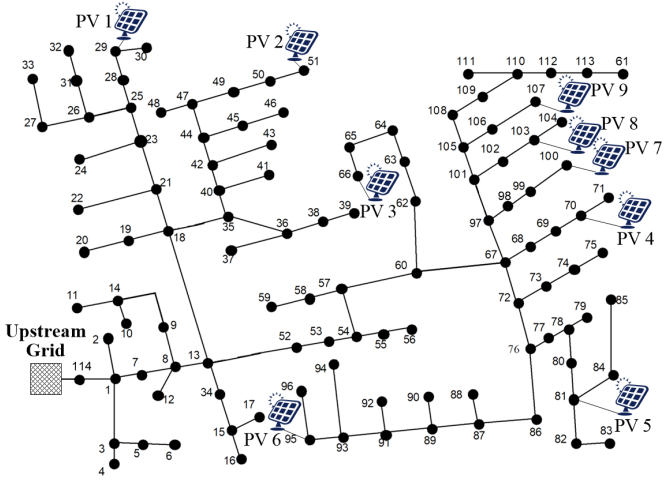

Fig. 6. IEEE 123-bus system augmented with 9 PV units.

TABLE I
REGULATION RANGE OF REACTIVE POWER

| Distribution System | Bus Location | Regulation Range (Mvar) |
|---|---|---|
| IEEE 33-bus | 16, 18, 31, 33 | [-0.45, 0.45] |
|  | 17, 32 | [-1, 1] |
| IEEE 123-bus | 29, 51, 66, 70, 81, 95, 100, 103, 107 | [-0.36, 0.36] |
|  | 28, 48, 65, 71, 84, 92, 104, 111 | [-0.6, 0.6] |

The proposed UVCP method is compared with the PPO method. Considering that power prediction results are affected by the selection of algorithms, we use three different neural networks, including the long short-term memory (LSTM) network, temporal convolutional network (TCN), and transformer (TF) network, to perform probabilistic prediction of renewable energy power generation and load demands. Therefore, the PPO methods are specified as PPO-LSTM, PPO-TCN, and PPO-TF, respectively. The neural networks are built using the Darts toolbox [33], [34], which is a cutting-edge Python library for time-series prediction. Table II gives the main hyperparameters of these neural networks, which are determined by grid search methods. Other hyperparameters are set as default ones of the Darts toolbox.

TABLE II
MAIN HYPERPARAMETERS OF NEURAL NETWORKS USED IN PPO METHOD

| Type | Hyperparameters |
|---|---|
| LSTM | hidden units: 300, training window length: 10, input chunk length: 3 |
| TCN | dilation base: 2, kernel size: 6, number of filters: 3, weight normalization: True, input chunk length: 7 |
| TF | number of attention heads: 4, embed dimension: 16, number of encoder layers: 2, number of decoder layers: 2, dimension of feedforward networks: 64, input chunk length: 3 |

The program of the UVCP method is developed in Matlab R2020a. The UVC prediction models are established based on the LSTM-based point predictions of power generation and loads. The optimization problems are constructed by Yalmip [35] and solved by Gurobi [36] on a laptop with Intel i7-10510U 1.8 GHz CPU and 16 GB RAM.

### B. Optimization Results

In this subsection, we respectively utilize the UVCP and PPO methods to solve the VaR-based voltage risk management problem to obtain reactive power generation strategies. Then, we analyze the nodal voltages of distribution systems under the obtained reactive power generation using the testing data of power generation and loads. We find that the voltage violation problem of the conventional PPO method between 10:00 and 15:00 exhibits similar characteristics, with the voltage risks for many buses exceeding the desired level of 0.05, defined as "over-limit" in this study. Therefore, we present the results for 13:00 as a representative interval when analyzing the single hourly interval. In addition, the voltage violation frequency of each bus at each hour is analyzed, and the corresponding results are given and discussed in the following section.

Figs. 7 and 8 show the voltage distributions of the 33-bus and 123-bus systems at 13:00 of each day. The medians of voltages obtained by different methods are similar, but the voltage violations are quite different. Only the UVCP method can ensure that voltage risks are smaller than or close to 0.05. Moreover, the UVCP method yields smaller voltage violations in extreme scenarios compared with other methods. This is because the UVCP method assesses the risk of voltage violations accurately.

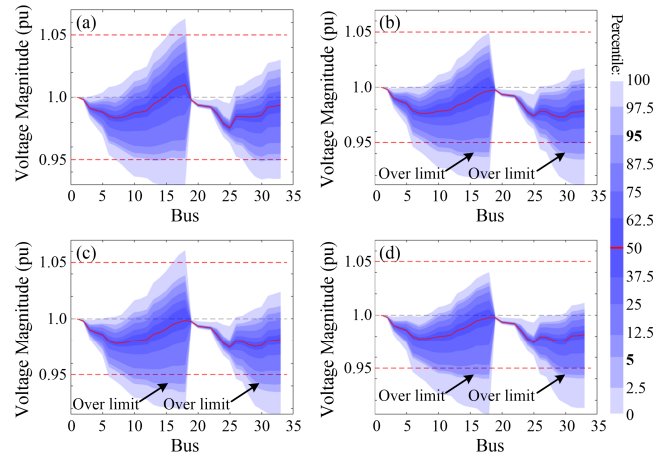

Fig. 7. Voltage distribution of the IEEE 33-bus system at 13:00 obtained by (a) UVCP, (b) PPO-LSTM, (c) PPO-TCN, and (d) PPO-TF, where the desired risk level of voltage violation is 0.05.

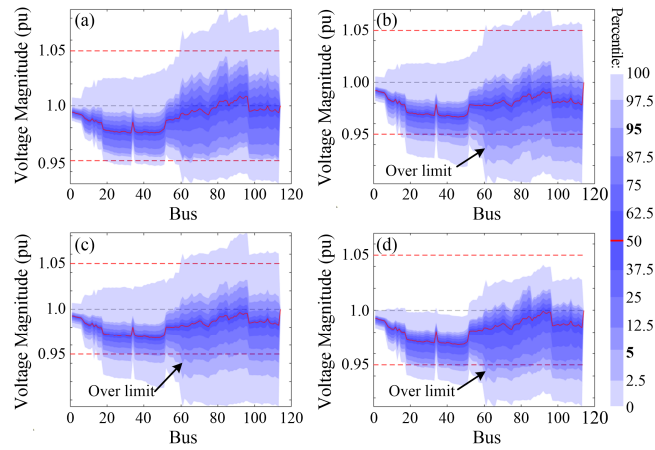

Fig. 8. Voltage distribution of the IEEE 123-bus system at 13:00 obtained by (a) UVCP, (b) PPO-LSTM, (c) PPO-TCN, and (d) PPO-TF, where the desired risk level of voltage violation is 0.05.

Figs. 9 (a) and (b) show the box diagrams of the costs of purchasing reactive power in the 33-bus and 123-bus systems, respectively. The cost medians of the UVCP method are higher than those of PPO methods. This is because the PPO methods underestimate voltage risks and the costs of purchasing reactive power are smaller than expected. Moreover, the differences in reactive power costs obtained by UVCP and PPO methods are small. Specifically, compared with the results obtained by the PPO methods, the increases in the average costs of purchasing reactive power obtained by the UVCP method are about 20% and 5% in the 33-bus and 123-bus systems, respectively. Therefore, the proposed UVCP method offers rational reactive power costs to guarantee the desired voltage risks.

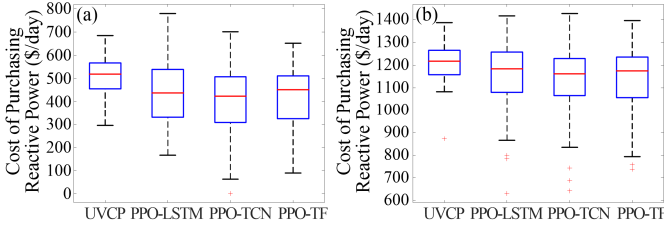

Fig. 9. Box diagrams of costs of purchasing reactive power: (a) IEEE 33-bus system, (b) IEEE 123-bus system.

### C. Voltage Risks

This subsection further gives voltage violation frequencies to compare the voltage risks under the reactive power generation strategies obtained by different methods. The voltage violation frequencies are calculated for each bus at each hour by the following formulas:

$$\begin{cases} \text{Freq}\{v_{i,t} > v_{i,\max}\} = \sum_{y \in \mathbb{Y}} 1(v_{i,t,y} > v_{i,\max})/N_{\mathbb{Y}} \\ \text{Freq}\{v_{i,t} < v_{i,\min}\} = \sum_{y \in \mathbb{Y}} 1(v_{i,t,y} < v_{i,\min})/N_{\mathbb{Y}} \end{cases} \quad (26)$$

where $1(\cdot)$ is the indicator function, $v_{i,t,y}$ is the squared voltage magnitude of bus $i$ at the $t_{\text{th}}$ hour of the $y_{\text{th}}$ day, $\mathbb{Y}$ is the day set of testing data, $N_{\mathbb{Y}}$ is the total days in $\mathbb{Y}$.

Table III gives the maximum voltage violation frequencies of all buses at all hours. The desired risk level of voltage violation is 0.05. For all the methods, the maximum frequencies of voltage violation regarding $v_{i,\max}$ are no larger than 0.05. However, the maximum frequencies of voltage violation regarding $v_{i,\min}$ are much larger than 0.05 for the PPO methods, while it is 0.064, which is slightly larger than 0.05, in the 33-bus system for the UVCP method.

TABLE III
MAXIMUM VOLTAGE VIOLATION FREQUENCY OF VOLTAGE RISK MANAGEMENT STRATEGIES OBTAINED BY DIFFERENT METHODS

| Distribution System | Method | Maximum Frequency of Voltage Violation | |
|---|---|---|---|
| | | $v_i > v_{i,\max}$ | $v_i < v_{i,\min}$ |
| IEEE 33-bus | UVCP | 0.027 (bus 18) | 0.064 (bus 32) |
| | PPO-LSTM | 0 (All buses) | 0.135 (bus 30) |
| | PPO-TCN | 0.005 (bus 16) | 0.130 (bus 30) |
| | PPO-TF | 0 (All buses) | 0.130 (bus 30) |
| IEEE 123-bus | UVCP | 0.011 (bus 81) | 0.041 (bus 75) |
| | PPO-LSTM | 0.014 (bus 92) | 0.151 (bus 65) |
| | PPO-TCN | 0.014 (bus 95) | 0.116 (bus 61) |
| | PPO-TF | 0.006 (bus 80) | 0.119 (bus 65) |

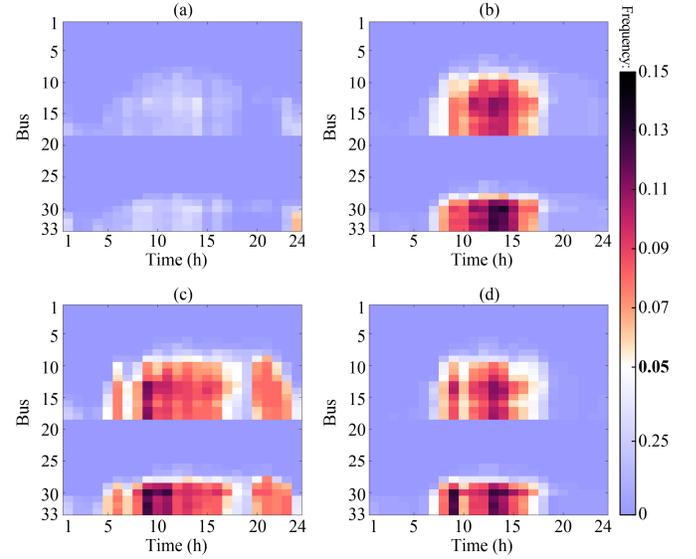

Fig. 10. Voltage violation frequency Freq$\{v_{i,t} < v_{i,\min}\}$ of each bus at each hour in the IEEE 33-bus system obtained by (a) UVCP, (b) PPO-LSTM, (c) PPO-TCN, and (d) PPO-TF.

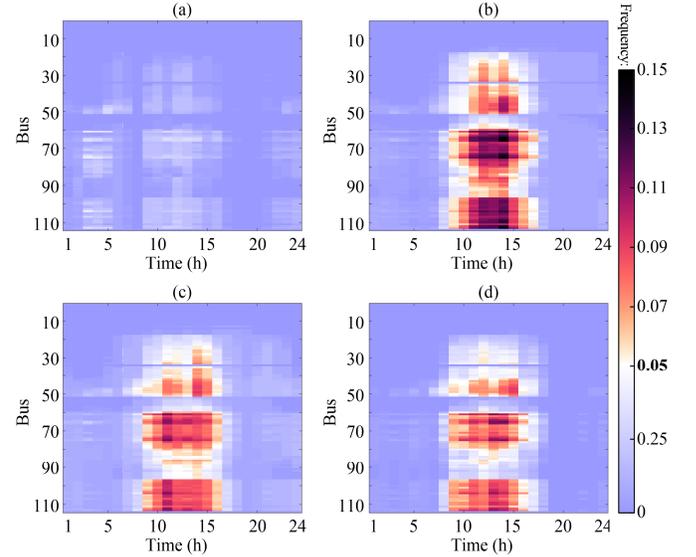

Fig. 11. Voltage violation frequency Freq$\{v_{i,t} < v_{i,\min}\}$ of each bus at each hour in the IEEE 123-bus system obtained by (a) UVCP, (b) PPO-LSTM, (c) PPO-TCN, and (d) PPO-TF.

Figs. 10 and 11 show the Freq$\{v_{i,t} < v_{i,\min}\}$ of 33-bus and 123-bus systems obtained by different methods, respectively. For the PPO methods, the Freq$\{v_{i,t} < v_{i,\min}\}$ of many buses in the two systems are much larger than the desired risk level of 0.05. The notably high frequency of voltage violations mainly occurs between 10:00 and 15:00 due to significant non-Gaussian uncertainties in the active power outputs of PV units. The voltage variations caused by these uncertainties cannot be accurately assessed using joint Gaussian distribution-based probabilistic prediction models of nodal power injections. In contrast, only the Freq$\{v_{i,t} < v_{i,\min}\}$ of buses 33 and 32 in the 33-bus system at 24:00 are slightly larger than 0.05 for the UVCP method. In addition, the Freq$\{v_{i,t} < v_{i,\min}\}$ of many buses are close to zero in all methods. This is because these buses are near the upstream grid and their voltages are not significantly affected by the variable nodal power. The Freq$\{v_{i,t} < v_{i,\min}\}$ of all the buses are also close to zero, when the power outputs of PV units are zero and the load demands are low.

Fig. 12 shows $\text{VaR}_{0.95}(v_{18}^{\text{r}})$ and $-\text{VaR}_{0.95}(-v_{18}^{\text{r}})$ in the IEEE 33-bus system at 13:00 of each day from Mar. 18, 2014 to Jun. 26, 2014. VaRs are calculated by (14) in the UVCP method, and they are obtained from the left side of constraints (A12a) and (A12b) in the PPO method. As shown in the figures, the actual value of $v_{18}^{\text{r}}$ is strongly fluctuant. $-\text{VaR}_{0.95}(-v_{18}^{\text{r}})$ obtained by PPO is much larger than that by UVCP, and it is also larger than actual values in many situations, which means that the PPO method estimates $-\text{VaR}_{0.95}(-v_{18}^{\text{r}})$ in a low confidence level. Meanwhile, $\text{VaR}_{0.95}(v_{18}^{\text{r}})$ obtained by PPO is larger than that by UVCP; thus, the PPO method provides a more conservative estimation of upper bounds. To analyze the accuracy of VaR estimation, the confidence levels of $\text{VaR}_\tau(v_{18}^{\text{r}})$ and $-\text{VaR}_\tau(-v_{18}^{\text{r}})$ are calculated by (27), and the results are given in Table IV.

$$\begin{cases} \tau_{\text{act},+} = \sum_{y \in \mathbb{Y}} 1(\text{VaR}_{\tau,\text{est}}(v_{18,y}^{\text{r}}) > v_{18,y}^{\text{r}})/N_\mathbb{Y} \\ \tau_{\text{act},-} = \sum_{y \in \mathbb{Y}} 1(-\text{VaR}_{\tau,\text{est}}(-v_{18,y}^{\text{r}}) < v_{18,y}^{\text{r}})/N_\mathbb{Y} \end{cases} \quad (27)$$

where $v_{18,y}^{\text{r}}$ is the voltage of bus 18 at 13:00 of the $y_{\text{th}}$ day, $\text{VaR}_{\tau,\text{est}}(v_{18,y}^{\text{r}})$ and $-\text{VaR}_{\tau,\text{est}}(-v_{18,y}^{\text{r}})$ are the estimated $\text{VaR}_\tau(v_{18}^{\text{r}})$ and $-\text{VaR}_\tau(-v_{18}^{\text{r}})$, respectively, and $\tau_{\text{act},+}$ and $\tau_{\text{act},-}$ are the confidence levels of $\text{VaR}_{\tau,\text{est}}(v_{18,y}^{\text{r}})$ and $-\text{VaR}_{\tau,\text{est}}(-v_{18,y}^{\text{r}})$, respectively.

For the PPO methods, the confidence levels of $\text{VaR}_\tau(v_{18}^{\text{r}})$ are too large, while the confidence levels of $-\text{VaR}_\tau(-v_{18}^{\text{r}})$ are too small. The small confidence levels of $-\text{VaR}_\tau(-v_{18}^{\text{r}})$ cause that Freq$\{v_{i,t} < v_{i,\min}\}$ obtained by the PPO methods is much larger than 0.05, as shown in Figs. 10 and 11. The variable weather conditions can cause substantial decreases in the power outputs of PV units. This risk is often underestimated in the PPO methods, leading to overly optimistic estimations of lower voltage bounds. The UVCP method provides more accurate voltage risk assessments because of predicting accurate probability models of UVCs, rather than predicting probability models of power generation and load demands. Hence, the voltage risk obtained by the UVCP method is consistent with the desired one.

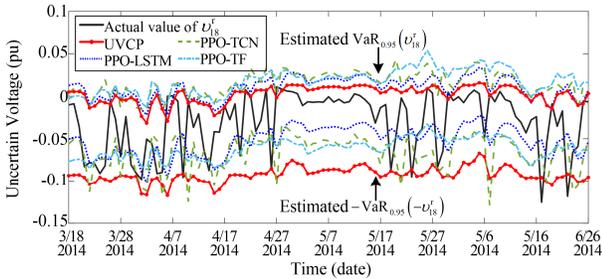

Fig. 12. $\text{VaR}_{0.95}(v_{18}^{\text{r}})$ and $-\text{VaR}_{0.95}(-v_{18}^{\text{r}})$ of the IEEE 33-bus system at 13:00 of each day from Mar. 18, 2014 to Jun. 26, 2014.

TABLE IV
CONFIDENCE LEVELS OF $\text{VaR}_\tau(v_{18}^{\text{r}})$ AND $-\text{VaR}_\tau(-v_{18}^{\text{r}})$ OF IEEE 33-BUS SYSTEM AT 13:00 OBTAINED BY DIFFERENT METHODS

| Risk index | Confidence Level $\tau$ | | | |
|---|---|---|---|---|
| | Method | Value | Method | Value |
| $\text{VaR}_\tau(v_{18}^{\text{r}})$ | Desired value | 0.950 | UVCP | 0.959 |
| | PPO-LSTM | 0.977 | PPO-TCN | 0.982 |
| | PPO-TF | 0.995 | - | - |
| $-\text{VaR}_\tau(-v_{18}^{\text{r}})$ | Desired value | 0.950 | UVCP | 0.966 |
| | PPO-LSTM | 0.838 | PPO-TCN | 0.881 |
| | PPO-TF | 0.859 | - | - |

### D. Comparison Between VaR and CVaR

Figs. 13 (a) and (b) show the maximum frequencies of voltage violation at each bus obtained by the UVCP method using VaR and CVaR constraints, respectively. The voltage violation frequencies under the CVaR constraints are about half of those under the VaR constraints. Figs. 14 (a) and (b) show the box diagrams of the costs of purchasing reactive power and the energy curtailment of PV units, respectively. The CVaR constraints yield a higher cost as well as more curtailed active power than the VaR constraints. This is because the CVaR is a conservative approximation of the VaR according to (12). In real-world applications, the VaR or CVaR could be selected based on users' preferences. Meanwhile, it should be noted that CVaR-based constraints would produce smaller risks while causing larger operating costs.

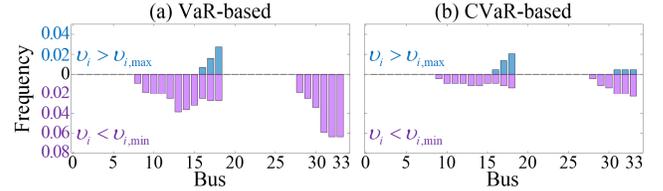

Fig. 13. Maximum voltage violation frequency of each bus of IEEE 33-bus system: (a) VaR-based results and (b) CVaR-based results.

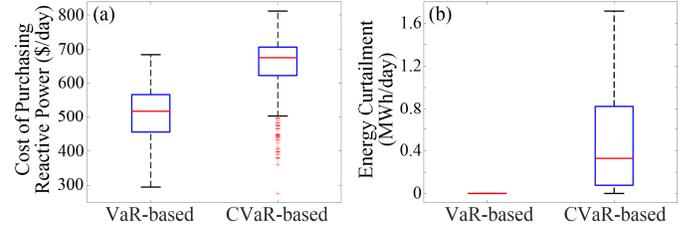

Fig. 14. Box diagrams of objective values of IEEE 33-bus system in VaR-based and CVaR-based results, where subgraph (a) is the cost of purchasing reactive power and subgraph (b) is the energy curtailment of PV units.

### E. Calculation Time

Table V shows the average calculation time of different methods in the two systems. When only reactive power generation is involved, both UVCP-based and PPO-based voltage risk management problems are LPs. When both reactive power generation and active power curtailment are utilized, the UVCP-based voltage risk management problem is cast as an MILP problem, and the PPO-based voltage risk management problem is transformed into a conic programming (CP) problem. As shown in the table, the calculation time of the proposed UVCP method is shorter than or equal to that of the PPO methods. The time to solve the optimization problems is increased when the active power curtailment of PV units is used. Overall, it is computationally efficient to obtain voltage risk management results.

TABLE V
AVERAGE CALCULATION TIME OF DIFFERENT METHODS

| Distribution System | Method | Active Power Curtailment | Problem Type | Time (s) |
|---|---|---|---|---|
| IEEE 33-bus | UVCP | No | LP | 0.06 |
| | | Yes | MILP | 0.14 |
| | PPO | No | LP | 0.06 |
| | | Yes | CP | 0.17 |
| IEEE 123-bus | UVCP | No | LP | 0.13 |
| | | Yes | MILP | 0.31 |
| | PPO | No | LP | 0.15 |
| | | Yes | CP | 0.41 |

## VI. Conclusion

This paper develops the UVC prediction method for voltage risk assessment and management in power distribution systems. First, UVCs are derived from nodal voltage magnitude. Then, the probabilistic UVC prediction method is developed, and the VaR and CVaR formulas of UVCs are established. Finally, the voltage risk management problem is reformulated into an LP or MILP problem to obtain reactive power generation and active power curtailment strategies.

The proposed method is tested on power distribution systems with PV units and is compared with the conventional method, in which the uncertain power generation and loads are modeled via probabilistic prediction. Numerical results show that the proposed method provides voltage risk management strategies that satisfy desired risks. The difference between the maximum voltage violation frequency and the desired risk is about 0.01, which is only about 15% of that of other methods. Moreover, the average calculation time of the proposed method is shorter than or equal to that of other methods.

The superior performance of the proposed method owes to the accurate assessment of UVCs under uncertain power generation and loads, instead of performing the challenging task of high-dimensional probabilistic prediction of power generation and loads. The proposed UVC prediction method can be applied to other risk assessments, and it will be utilized to solve other stochastic problems in power system operations.

## Appendix

### A. Derivation of $\text{VaR}_\tau(v_i^r)$

According to the VaR definition, the following equation is established for $\text{VaR}_\tau(v_i^r | \tilde{v}_i^r)$:

$$\int_{-\infty}^{\text{VaR}_\tau(v_i^r | \tilde{v}_i^r)} f_{v_i^r | \tilde{v}_i^r}(x) dx = \tau \quad (A1)$$

Because $v_i = v_i^r + v_i^c + v_i^o$, the relationship between the PDFs of $v_i^r$ and $v_i$ is expressed as:

$$f_{v_i^r | \tilde{v}_i^r}(x) = f_{v_i}(x + v_i^c + v_i^o) \quad (A2)$$

Let $y = x + v_i^c + v_i^o$, and by substituting (A2) into (A1), we obtain:

$$\int_{-\infty}^{\text{VaR}_\tau(v_i^r | \tilde{v}_i^r) + v_i^c + v_i^o} f_{v_i}(y) dy = \tau \quad (A3)$$

Equation (A3) indicates that $\text{VaR}_\tau(v_i) = \text{VaR}_\tau(v_i^r | \tilde{v}_i^r) + v_i^c + v_i^o$.

### B. Newton-Raphson Method to Calculate $\text{VaR}_\tau(v_i^r | \tilde{v}_i^r)$

The iteration formula of the Newton-Raphson method to solve (14) is given as:

$$v_i^{r(w+1)} = v_i^{r(w)} + \left[\tau - F\left(v_i^{r(w)} | \tilde{v}_i^r\right)\right] / f\left(v_i^{r(w)} | \tilde{v}_i^r\right) \quad (A4)$$

where $v_i^{r(w)}$ and $v_i^{r(w+1)}$ are the values of $\text{VaR}_\tau(v_i^r | \tilde{v}_i^r)$ in the $w_{\text{th}}$ and $(w+1)_{\text{th}}$ iterations, respectively. The initial value $v_i^{r(0)}$ is given by:

$$v_i^{r(0)} = \text{E}\left(v_i^r | \tilde{v}_i^r\right) + \Phi^{-1}(\tau)\text{D}\left(v_i^r | \tilde{v}_i^r\right) \quad (A5)$$

$$\begin{cases} \text{E}\left(v_i^r | \tilde{v}_i^r\right) = \sum_{k=1}^{K} \omega_{i,k}^* \mu_{i,k}^* \\ \text{D}\left(v_i^r | \tilde{v}_i^r\right) = \sum_{k=1}^{K} \omega_{i,k}^* \left(\Sigma_{i,k}^* + \mu_{i,k}^{*2}\right) - \left(\sum_{k=1}^{K} \omega_{i,k}^* \mu_{i,k}^*\right)^2 \end{cases} \quad (A6)$$

where $\Sigma_{i,k}^*$, $\mu_{i,k}^*$, and $\omega_{i,k}^*$ are the variance, mean, and weight coefficient of the $k_{\text{th}}$ component of the conditional PDF (10), respectively, $\Phi^{-1}(\cdot)$ represents the inverse of the standard normal cumulative distribution function, and $\text{E}(\cdot)$ and $\text{D}(\cdot)$ are mean and variance operators, respectively.

### C. Derivation of $\text{CVaR}_\tau(v_i^r)$

According to the CVaR definition, the following equation is established for $\text{CVaR}_\tau(v_i^r | \tilde{v}_i^r)$:

$$\text{CVaR}_\tau(v_i^r | \tilde{v}_i^r) = \frac{1}{1-\tau} \int_{\text{VaR}_\tau(v_i^r | \tilde{v}_i^r)}^{+\infty} x f_{v_i^r | \tilde{v}_i^r}(x) dx \quad (A7)$$

Let $y = x + v_i^c + v_i^o$, and by substituting (A2) into (A7), we obtain:

$$\begin{aligned}
\text{CVaR}_\tau\left(v_i^r | \tilde{v}_i^r\right) &= \frac{1}{1-\tau} \int_{\text{VaR}_\tau(v_i^r | \tilde{v}_i^r) + v_i^c + v_i^o}^{+\infty} (y - v_i^c - v_i^o) f_{v_i}(y) dy \\
&= \frac{1}{1-\tau} \int_{\text{VaR}_\tau(v_i^r | \tilde{v}_i^r) + v_i^c + v_i^o}^{+\infty} y f_{v_i}(y) dy - \frac{(v_i^c + v_i^o)}{1-\tau} \int_{\text{VaR}_\tau(v_i^r | \tilde{v}_i^r) + v_i^c + v_i^o}^{+\infty} f_{v_i}(y) dy \\
&= \frac{1}{1-\tau} \int_{\text{VaR}_\tau(v_i)}^{+\infty} y f_{v_i}(y) dy - \frac{(v_i^c + v_i^o)}{1-\tau} \int_{\text{VaR}_\tau(v_i^r | \tilde{v}_i^r)}^{+\infty} f_{v_i^r | \tilde{v}_i^r}(x) dx \\
&= \text{CVaR}_\tau(v_i) - (v_i^c + v_i^o)
\end{aligned} \quad (A8)$$

Hence, $\text{CVaR}_\tau(v_i) = \text{CVaR}_\tau(v_i^r | \tilde{v}_i^r) + v_i^c + v_i^o$.

### D. Closed-form Expression to Calculate $\text{CVaR}_\tau(v_i^r | \tilde{v}_i^r)$

According to the CVaR definition and (10), $\text{CVaR}_\tau(v_i^r | \tilde{v}_i^r)$ is expressed as:

$$\begin{aligned}
\text{CVaR}_\tau\left(v_i^r | \tilde{v}_i^r\right) &= \frac{1}{1-\tau} \int_{\text{VaR}_\tau(v_i^r | \tilde{v}_i^r)}^{+\infty} x f(x) dx \\
&= \frac{1}{1-\tau} \sum_{k=1}^{K} \omega_{i,k}^* \left( \int_{\text{VaR}_\tau(v_i^r | \tilde{v}_i^r)}^{+\infty} x f_k(x) dx \right)
\end{aligned} \quad (A9)$$

Because $f_k(\cdot)$ is the Gaussian distribution function, the integration $\int_{\text{VaR}_\tau(v_i^r | \tilde{v}_i^r)}^{+\infty} x f_k(x) dx$ is obtained as:

$$\int_{\text{VaR}_\tau(v_i^r | \tilde{v}_i^r)}^{+\infty} x f_k(x) dx = \begin{pmatrix} \mu_{i,k}^* \left\{1 - F_k\left[\text{VaR}_\tau(v_i^r | \tilde{v}_i^r)\right]\right\} \\ + \Sigma_{i,k}^* f_k\left[\text{VaR}_\tau(v_i^r | \tilde{v}_i^r)\right] \end{pmatrix} \quad (A10)$$

By substituting (A10) into (A9), the closed-form expression to calculate $\text{CVaR}_\tau(v_i^r | \tilde{v}_i^r)$ is derived, as given by (16).

### E. Optimization Model of PPO Method

The predicted PDF of power generation and loads is given as:

$$f\left(\begin{bmatrix} \chi \\ \zeta \end{bmatrix}\right) = \text{N}\left(\begin{bmatrix} \chi \\ \zeta \end{bmatrix} \begin{bmatrix} \mu_\chi \\ \mu_\zeta \end{bmatrix}, \begin{bmatrix} \Sigma_\chi & \Sigma_{\chi\zeta} \\ \Sigma_{\zeta\chi} & \Sigma_\zeta \end{bmatrix}\right) \quad (A11)$$

where $\mu$ and $\Sigma$ are the mean vector and covariance matrix, and $\text{N}(\cdot)$ is the Gaussian distribution function.

Based on (A11), the VaR constraint (18) is reformulated as the following conic constraints [9], [10]:

$$\text{E}(v_i) + \Phi^{-1}(\tau)\sqrt{\text{D}(v_i)} \leq v_{i,\max} \quad (A12a)$$

$$\mathrm{E}(\upsilon_i) - \Phi^{-1}(\tau)\sqrt{\mathrm{D}(\upsilon_i)} \geq \upsilon_{i,\min} \quad \text{(A12b)}$$

$$\mathrm{E}(\upsilon_i) = (1-\alpha)\boldsymbol{b}_{i,\mathbb{G}}^{\mathrm{T}}\boldsymbol{\mu}_{\chi} - \boldsymbol{b}_{i,\mathbb{D}}^{\mathrm{T}}\boldsymbol{\mu}_{\zeta} + \upsilon_i^{\mathrm{c}} + \upsilon_i^{\mathrm{o}} \quad \text{(A12c)}$$

$$\mathrm{D}(\upsilon_i) = (1-\alpha)^2 \boldsymbol{b}_{i,\mathbb{G}}^{\mathrm{T}}\boldsymbol{\Sigma}_{\chi}\boldsymbol{b}_{i,\mathbb{G}} + \boldsymbol{b}_{i,\mathbb{D}}^{\mathrm{T}}\boldsymbol{\Sigma}_{\zeta}\boldsymbol{b}_{i,\mathbb{D}} - 2(1-\alpha)\boldsymbol{b}_{i,\mathbb{G}}^{\mathrm{T}}\boldsymbol{\Sigma}_{\chi\zeta}\boldsymbol{b}_{i,\mathbb{D}} \quad \text{(A12d)}$$

where $\boldsymbol{b}_{i,\mathbb{G}}$ and $\boldsymbol{b}_{i,\mathbb{D}}$ are the vectors of $b_{i,g}$ and $b_{i,d}$, respectively.

## REFERENCES

[1] P. Li, B. Jin, D. Wang, and B. Zhang, "Distribution system voltage control under uncertainties using tractable chance constraints," *IEEE Trans. Power Syst.*, vol. 34, no. 6, pp. 5208–5216, Nov. 2019, doi: 10.1109/TPWRS.2018.2880406.

[2] B. Xu, G. Zhang, K. Li, B. li, H. Chi, Y. Yao, and Z. Fan, "Reactive power optimization of a distribution network with high-penetration of wind and solar renewable energy and electric vehicles," *Prot. Control Mod. Power Syst.*, vol. 7, no. 51, pp. 1–13, Dec. 2022, doi: 10.1186/s41601-022-00271-w.

[3] Y. Zhang, K. Lin, W. Deng, D. Zhang and D. Xiao, "A Flexible Voltage Control Strategy Based on Stage-Division for Microgrids," *Prot. Control Mod. Power Syst.*, vol. 9, no. 3, pp. 60-69, May 2024, doi: 10.23919/PCMP.2023.000137.

[4] C. Wang, M. Li, Y. Cao, et al. "Gradient boosting dendritic network for ultra-short-term PV power prediction," *Front. Energy*, early access, 2024, doi: 10.1007/s11708-024-0915-y.

[5] C. Xia, W. Li, X. Chang, T. Yang, and A. Y. Zomaya, "A rank-based multiple-choice secretary algorithm for minimising microgrid operating cost under uncertainties," *Front. Energy*, vol. 17, no. 2, pp. 198–210, Apr. 2023, doi: 10.1007/s11708-023-0874-8.

[6] X. Geng and L. Xie, "Data-driven decision making in power systems with probabilistic guarantees: Theory and applications of chance constrained optimization," *Annu. Rev. Control*, vol. 47, pp. 341–363, May 2019, doi: 10.1016/j.arcontrol.2019.05.005.

[7] F. Ul Nazir, B. C. Pal, and R. A. Jabr, "A two-stage chance constrained volt/var control scheme for active distribution networks with nodal power uncertainties," *IEEE Trans. Power Syst.*, vol. 34, no. 1, pp. 314–325, Jan. 2019, doi: 10.1109/TPWRS.2018.2859759.

[8] K. S. Ayyagari, R. Gonzalez, Y. Jin, M. Alamaniotis, S. Ahmed, and N. Gatsis, "Learning reactive power control polices in distribution networks using conditional value-at-risk and artificial neural networks," *J. Mod. Power Syst. Clean Energy*, vol. 11, no. 1, pp. 201–211, Jan. 2023, doi: 10.35833/MPCE.2022.000477.

[9] D. Bienstock, M. Chertkov, and S. Harnett, "Chance-constrained optimal power flow: Risk-aware network control under uncertainty," *Siam Review*, vol. 56, no. 3, pp. 461–495, Aug. 2014, doi: 10.1137/130910312.

[10] D. Mak and D. H. Choi, "Optimization framework for coordinated operation of home energy management system and volt-var optimization in unbalanced active distribution networks considering uncertainties," *Appl. Energy*, vol. 276, pp. 115495, Jul. 2020, doi: 10.1016/j.apenergy.2020.115495.

[11] S. Najafi, M. M. Lakouraj, S. A. Sedgh, H. Livani, M. Benidris, and M. S. Fadali, "Chance-constraint volt-var optimization in PV-penetrated distribution networks," in *Proc. IEEE Kans. Power Energy Conf.*, Manhattan, KS, USA, pp. 1–6, 2022, doi: 10.1109/KPEC54747.2022.9814811.

[12] M. Mansourlakouraj, M. Gautam, H. Livani, M. Benidris, and P. Fajri, "Multi-timescale risk-constrained volt/var control of distribution grids with electric vehicles and solar inverters," in *Proc. IEEE PES Innov. Smart Grid Technol. Eur.*, Espoo, Finland, 2021, pp. 1–6, doi: 10.1109/ISGTEurope52324.2021.9640047.

[13] D. Ke, C. Y. Chung and Y. Sun, "A novel probabilistic optimal power flow model with uncertain wind power generation described by customized gaussian mixture model," *IEEE Trans. Sustain. Energy*, vol. 7, no. 1, pp. 200–212, Jan. 2016, doi: 10.1109/TSTE.2015.2489201.

[14] Y. Yang, W. Wu, B. Wang and M. Li, "Analytical reformulation for stochastic unit commitment considering wind power uncertainty with Gaussian mixture model," *IEEE Trans. Power Syst.*, vol. 35, no. 4, pp. 2769–2782, Jul. 2020, doi: 10.1109/TPWRS.2019.2960389.

[15] T. Chen, Y. Song, D. J. Hill, and A. Y. S. Lam, "Chance-constrained OPF in droop-controlled microgrids with power flow routers," *IEEE Trans. Smart Grid*, vol. 13, no. 4, pp. 2601–2613, Jul. 2022, doi: 10.1109/TSG.2022.3154151.

[16] B. Axel, "Mixture density networks for distribution and uncertainty estimation," M.S. thesis, Universitat de Barcelona, Barcelona, Spain, 2017.

[17] H. Zhang, Y. Liu, J. Yan, S. Han, L. Li and Q. Long, "Improved deep mixture density network for regional wind power probabilistic forecasting," *IEEE Trans. Power Syst.*, vol. 35, no. 4, pp. 2549–2560, Jul. 2020, doi: 10.1109/TPWRS.2020.2971607.

[18] D. Salinas, M. Bohlke-Schneider, L. Callot, R. Medico, and J. Gasthaus, "High-dimensional multivariate forecasting with low-rank Gaussian copula processes," in *Proc. 33rd conf. Adv. Neural Inf. Process. Syst.*, Vancouver, Canada, 2019, pp. 1–11, doi: 10.5555/3454287.3454900.

[19] N. Zhou, X. Xu, Z. Yan, and M. Shahidehpour, "Spatio-temporal probabilistic forecasting of photovoltaic power based on monotone broad learning system and copula theory," *IEEE Trans. Sustain. Energy*, vol. 13, no. 4, pp. 1874–1885, Oct. 2022, doi: 10.1109/TSTE.2022.3174012.

[20] M. Baran and F. W. Wu, "Optimal capacitor placement on radial distribution systems," *IEEE Trans. Power Del.*, vol. 4, no. 1, pp. 725–734, Jan. 1989, doi: 10.1109/61.19265.

[21] Q. Hou, G. Chen, N. Dai and H. Zhang, "Distributionally robust chance-constrained optimization for soft open points operation in active distribution networks," *CSEE J. Power Energy Syst.*, early access, 2023, doi: 10.17775/CSEEJPES.2021.02110.

[22] Q. Zhang, F. Bu, Y. Guo and Z. Wang, "Tractable data enriched distributionally robust chance-constrained conservation voltage reduction," *IEEE Trans. Power Syst.*, vol. 39, no. 1, pp. 821-835, 2024, doi: 10.1109/TPWRS.2023.3244895.

[23] Y. Gao, X. Xu, Z. Yan and M. Shahidehpour, "Gaussian mixture model for multivariate wind power based on kernel density estimation and component number reduction," *IEEE Trans. Sustain. Energy*, vol. 13, no. 3, pp. 1853–1856, Jul. 2022, doi: 10.1109/TSTE.2022.3159391.

[24] T. Jiang, X. Dong, R. Zhang, et al. "Active-reactive power scheduling of integrated electricity-gas network with multi-microgrids," *Front. Energy* vol. 17, no.2, pp. 251–265, Apr. 2023, doi: 10.1007/s11708-022-0857-1.

[25] V. Kekatos, G. Wang, A. J. Conejo, and G. B. Giannakis, "Stochastic reactive power management in microgrids with renewables," *IEEE Trans. Power Syst.*, vol. 30, no. 6, pp. 3386–3395, Nov. 2015, doi: 10.1109/TPWRS.2014.2369452.

[26] G. Chen, H. Zhang and Y. Song, "Chance-constrained DC optimal power flow with non-Gaussian distributed uncertainties," in *Proc. IEEE Power Energy Soc. Gen. Meet.*, Denver, CO, USA, 2022, pp. 1–5, doi: 10.1109/PESGM48719.2022.9916658.

[27] W. Wei and J. Wang, "Modeling and Optimization of Interdependent Energy Infrastructures," Berlin, Germany: Springer, 2020.

[28] S. K. Goswami and S. K. Basu, "A new algorithm for the reconfiguration of distribution feeders for loss minimization," *IEEE Trans. Power Del.*, vol. 7, no. 3, pp. 1484–1491, Jul. 1992, doi: 10.1109/61.141868.

[29] L. Bobo, A. Venzke, and S. Chatzivasileiadis, "Second-order cone relaxations of the optimal power flow for active distribution grids: Comparison of methods," *Int. J. Electr. Power Energy Syst.*, vol. 127, 106625, May 2021, doi: 10.1016/j.ijepes.2020.106625.

[30] [Online]. Available: https://data.solaranywhere.com/Data/Public.

[31] Trindade, Artur. ElectricityLoadDiagrams20112014. *UCI Machine Learning Repository*, 2015. doi: 10.24432/C58C86.

[32] [Online]. Available: https://github.com/matrixEE/UVQCaseData.

[33] J. Herzen et al., "Darts: User-Friendly Modern Machine Learning for Time Series," *J. Mach. Learn. Res.*, vol. 23, no. 1, pp. 5442–5447, Jan. 2022.

[34] [Online]. Available: https://unit8co.github.io/darts/.

[35] J. Löfberg, "YALMIP: A toolbox for modeling and optimization in MATLAB," in *Proc. IEEE Int. Symp. Comput. Aided Control Syst. Des.*, New Orleans, LA, USA, 2004, pp. 284–289, doi: 10.1109/CACSD.2004.1393890.

[36] Gurobi Optimization, LLC, "Gurobi Optimizer Reference Manual." 2022, [Online]. Available: https://www.gurobi.com.